\documentclass[runningheads]{llncs}

\usepackage{accv}       

\usepackage{accvabbrv} 
\usepackage{graphicx}
\usepackage{booktabs}
\usepackage{multirow}
\usepackage{multicol}
\usepackage[normalem]{ulem}
\usepackage[accsupp]{axessibility}  
\usepackage[breaklinks,colorlinks,citecolor=accvblue]{hyperref}
\usepackage{hyperref}   
\usepackage{orcidlink}  
\makeatletter
\newcommand{\printfnsymbol}[1]{%
  \textsuperscript{\@fnsymbol{#1}}%
}
\makeatother

\begin{document}

\title{Nash Meets Wertheimer:\\ Using Good Continuation in Jigsaw Puzzles} 

\titlerunning{Nash Meets Wertheimer: Using Good Continuation in Jigsaw Puzzles}

\author{{Marina Khoroshiltseva\thanks{Equal contribution.\\
E-mails: \email{m.khoroshiltseva@unive.it, luca.palmieri@unive.it, sinem.aslan@unive.it, 
sebastiano.vascon@unive.it, pelillo@unive.it}}\textsuperscript{,}\inst{1}\orcidlink{0000-0003-0424-0661} \and
Luca Palmieri\printfnsymbol{1}\textsuperscript{,}\inst{1}\orcidlink{0000-0002-9701-1915} \and
Sinem Aslan\inst{1,2}\orcidlink{0000-0003-0068-6551} \and
Sebastiano Vascon\inst{1,2}\orcidlink{0000-0002-7855-1641} \and
Marcello Pelillo\inst{1}\orcidlink{0000-0001-8992-9243}}}

\authorrunning{M. Khoroshiltseva et al.}

\institute{Department of Computer Science (DAIS), Ca’ Foscari University of Venice, Italy 
\and European Center for Living Technology (ECLT), Venice, Italy
}

\maketitle
\begin{abstract}
Jigsaw puzzle solving is a challenging task for computer vision since it requires high-level spatial and semantic reasoning. To solve the problem, existing approaches invariably use color and/or shape information but in many real-world scenarios, such as in archaeological fresco reconstruction, this kind of clues is often unreliable due to severe physical and pictorial deterioration of the individual fragments. This makes state-of-the-art approaches entirely unusable in practice. On the other hand, in such cases, simple geometrical patterns such as lines or curves offer a powerful yet unexplored clue. In an attempt to fill in this gap, in this paper we introduce a new challenging version of the puzzle solving problem in which one deliberately ignores conventional color and shape features and relies solely on the presence of linear geometrical patterns. The reconstruction process is then only driven by one of the most fundamental principles of Gestalt perceptual organization, namely Wertheimer's {\em law of good continuation}. In order to tackle this problem, we formulate the puzzle solving problem as the problem of finding a Nash equilibrium of a (noncooperative) multiplayer game and use classical multi-population replicator dynamics to solve it. The proposed approach is general and allows us to deal with pieces of arbitrary shape, size and orientation. We evaluate our approach on both synthetic and real-world data and compare it with state-of-the-art algorithms. The results show the intrinsic complexity of our purely line-based puzzle problem as well as the relative effectiveness of our game-theoretic formulation.
\end{abstract}

\section{Introduction}
\label{sec:intro}

The problem of automatically solving jigsaw puzzles has attracted the interest of computer scientists since the early 1960's \cite{FreGar64} and despite its intrinsic difficulty (the problem is known to be NP-hard \cite{demaine2007jigsaw}) it still remains a subject of intense studies within the computer vision and pattern recognition communities \cite{harel2021crossing,gallagher2012jigsaw,son2018solving}.
This is because of its applications in such diverse domains as archaeology \cite{brown2008system, DERECH2021108065}, genome biology \cite{marande2007mitochondrial}, forensics \cite{deever2012semi}, speech recognition \cite{zhao2007puzzle}, etc.

\begin{figure}[t]
    \centering
    \includegraphics[width=1\textwidth]{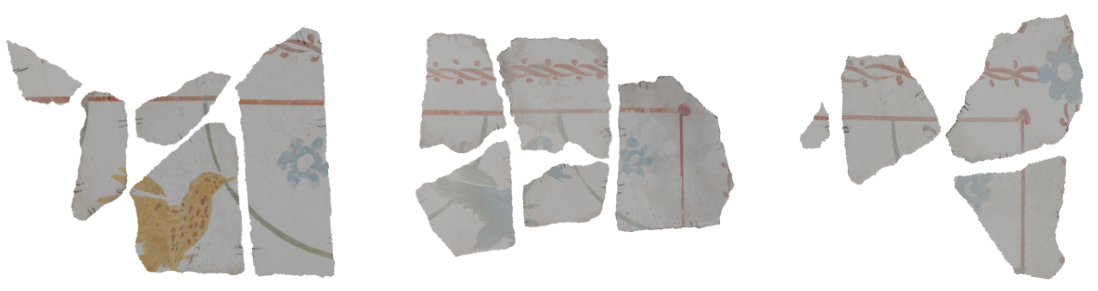}
    \caption{Three groups of fragments from an ancient broken fresco. Here, the good continuation principle plays a major role in the reassembly process.
    \vspace{-0.1cm}
}
    \label{fig:cover_story}
\end{figure}

Notwithstanding the tremendous progress made in the past decades, there still remains a large gap between the capabilities of lab-developed algorithms and the difficulties posed by real-world applications. One of the main reasons behind this state of affairs is that almost invariably existing algorithms rely only on color and/or shape information. While these clues work well on ``toy'' or commercial jigsaw puzzles, they often play a marginal role in real-world scenarios where individual pieces are severely deteriorated.
To illustrate this point, Fig. \ref{fig:cover_story} shows a few groups of fragments from an ancient shattered fresco. 
It is clear that here shape- and color-based compatibility functions are of little or no use in solving the puzzles. Nevertheless, we (humans) are able to effortlessly determine the relative locations of the fragments by simply ``matching'' the lines and curves depicted on them. The underlying principle behind this matching process is none other that Gestalt's law of good continuation originally proposed by Wertheimer \cite{Wer23} which, besides being one of the most fundamental laws of visual perception \cite{Kof63,RocPal90,Wag+12}, has also served as a motivation for countless computer vision algorithms \cite{SarBoy94,good_continuation,Pas+24}.

The situation described above is in fact quite common in the wild and arises also in contexts outside the archaeological domain every time one wants to automatically reassemble shattered physical artifacts (think, e.g., of the problem of reconstructing shredded documents containing engineering or architectural drawings). We maintain that, in order to bridge the gap between the lab and the real world, puzzle-solving algorithms should make a more extensive use of this kind of clues. 

\begin{figure}[t]
    \centering
    \includegraphics[width=\textwidth]{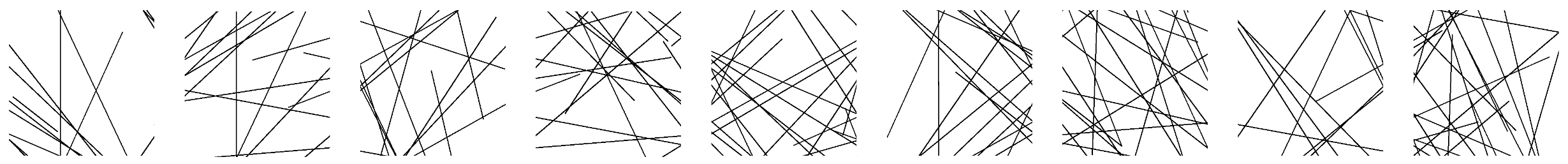}
    \vspace{-0.6cm}
    
    \caption{The difficulty of solving purely line-based jigsaw puzzles. The reader is invited to solve this small ($3 \times 3$) instance of the problem. Our algorithm solves it perfectly.
    \vspace{-0.4cm}
    }
    \label{fig:9_lines}
\end{figure}

The work described in this paper is a first step towards this direction. To explore the potential of purely line-based compatibility functions we introduce a new (abstract) version of the jigsaw puzzle problem in which we deliberately ignore information coming from the physical shape of the fragments as well as their color and rely instead exclusively on the presence of linear geometrical patterns. In so doing, we aim to explore the fundamental limitations of puzzle-solving algorithms in the absence of reliable color and shape information. 
Fig. \ref{fig:9_lines} shows a simple instance of our {\em line-based jigsaw puzzle problem}. As can be seen, despite its small size, the problem is far from simple and, not surprisingly, existing algorithms are (by construction) totally inadequate to solve it.

In order to attack the problem, we develop a novel game-theoretic approach which allows one to deal with pieces of arbitrary shape, size and orientation. We formulate the general jigsaw puzzle problem as a noncooperative multiplayer game where the players are the fragments, the (pure) strategies are the planar positions and orientations, and the payoffs reflect line-based compatibilities of fragments obtained using good continuation. Under this framework it is easy to see that puzzle solutions correspond to Nash equilibria of this ``jigsaw puzzle game'' \cite{nash1951non,OsbRub94} and in order to find them we use classical multi-population replicator dynamics from evolutionary game theory \cite{HofSig98,weibull1997evolutionary}.

We evaluate our approach on both synthetic and real-world datasets constructed specifically for this work and compare it with state-of-the-art algorithms. The experimental results show the intrinsic complexity of our purely line-based puzzle problem, confirm the inadequacy of existing approaches, and demonstrate the relative effectiveness of our game-theoretic formulation.

\section{Related Works}
\label{sec:related}
\vspace{-0.2cm}

The literature on puzzle-solving has predominantly focused on square jigsaw puzzles, where only pictorial information is available due to the identical shapes of all pieces, e.g., \cite{pomeranz2011fully, gallagher2012jigsaw, cho2010probabilistic, andalo2016psqp, son2018solving, paumard2020deepzzle, khoroshiltseva2021jigsaw, song2023siamese}. 
Earlier studies such as \cite{cho2010probabilistic, pomeranz2011fully} tackled square jigsaw puzzles with known piece size and orientation. Cho et al. \cite{cho2010probabilistic} used loopy belief propagation for graph-based piece arrangement, requiring low-resolution image estimations. Pomeranz et al. \cite{pomeranz2011fully} introduced a prediction-based dissimilarity metric and best-buddy concept for fully automatic solutions. Gallagher et al. \cite{gallagher2012jigsaw} addressed puzzles with unknown piece orientation, introducing a tree-based algorithm and a compatibility metric based on Mahalanobis distance. Paikin and Tal \cite{paikin2015solving} improved accuracy by modifying the dissimilarity metric of Pomeranz et al. \cite{pomeranz2011fully} and strategically selecting an anchor piece. Son et al. \cite{son2018solving} dealt with noise in pictorial content, and unknown piece orientation and puzzle size by requiring compatibility consensus within puzzle piece loops, extending beyond immediate neighbors. Besides these, a few works addressed solving puzzles with irregularly shaped fragments \cite{harel2021crossing, DERECH2021108065}, which hold potential applications in real-world scenarios such as archaeological puzzle solving. Derech and Tal \cite{DERECH2021108065} addressed archaeological puzzles in cultural heritage, characterized by irregularly shaped pieces prone to color fading, abrasion, and discontinuity, along with degraded fragment boundaries. They suggested pre-extrapolating each fragment and introduced confidence in dissimilarity scores, considering the uniqueness of match and fragment size during placement. Harel and Ben-Shahar \cite{harel2021crossing} introduced a new synthetically created puzzle type, namely the Crossing Cuts puzzle, and abstracted the problem as a multi-body spring-mass dynamical system supported with hierarchical loop constraints.

While certain works, e.g., \cite{pomeranz2011fully, paikin2015solving, son2018solving, harel2021crossing, DERECH2021108065}, have reported effective performances, they all employed greedy approaches, making decisions based only on local matchings without considering global consequences. Andalo et al. \cite{andalo2016psqp} was the first to formulate the problem as a global optimization problem, utilizing a Constrained Gradient Ascent algorithm for its solution. Khoroshiltseva et al. \cite{khoroshiltseva2021jigsaw} adopted a nonlinear relaxation labeling (RL) approach, extending the labeling update rule by Rosenfeld et al. \cite{rosenfeld1976scene}, encouraging but without guaranteeing feasible solutions. Vardi et al. \cite{VardiTKPB23} proposed a multi-phase approach for relaxation labeling, ensuring convergence to feasible puzzle solutions, even for puzzles with both known and unknown piece orientation. All these works adopting global optimization \cite{andalo2016psqp, khoroshiltseva2021jigsaw, VardiTKPB23} were designed specifically for the squared jigsaw puzzles with known puzzle sizes; with the assumption of known piece orientation in \cite{andalo2016psqp, khoroshiltseva2021jigsaw}.

In the last years, there has been an increased interest in neural network methods for puzzle solving. A common issue with learning-based approaches is that many struggle with scalability, performing well only on small-scale puzzles.
For instance, Paumard et al. \cite{paumard2020deepzzle} proposed a CNN model to predict neighboring relationships in 9-piece puzzles. Song et al. \cite{song2023siamese} introduced the Siamese-Discriminant Deep Reinforcement Learning method for 9-piece puzzles with eroded boundaries. More recently, diffusion models have begun to be utilized in puzzle-solving. These models involve introducing time-dependent noise into each piece's translation and rotation during the forward process, followed by iterative denoising during the backward process to reconstruct the original image. Hosseini et al. \cite{hossieni2024puzzlefusion} introduced PuzzleFusion for puzzles with up to 16 irregular-shaped fragments. Scarpellini et al. \cite{scarpellini2024diffassemble} represented puzzles as graphs and processed them with an Attention-based Graph Neural Network within a diffusion model framework. However, accuracy notably decreased for puzzles with 144 pieces compared to those with 36 pieces.

The choice of compatibility measurement significantly impacts puzzle-solving performance. While square jigsaw puzzles typically compare pictorial information along piece boundaries, irregularly shaped puzzles also consider matching fragment shapes. Various approaches exist: some compute color differences along fragment boundaries \cite{cho2010probabilistic, pomeranz2011fully, paikin2015solving, andalo2016psqp}, while others consider distributions of local gradients \cite{gallagher2012jigsaw, son2018solving}. Some works compare the length of matching boundaries between fragments to assess their shapes \cite{harel2021crossing}, while others combine all these measurements \cite{DERECH2021108065}. This study is the first, to our knowledge, to specifically address purely line-based compatibilities.

\section{Jigsaw Puzzles and Nash Equilibria}
\label{ssec:gametheory}
\vspace{-0.2cm}

\subsection{Nash equilibria}
We start this section by introducing some fundamental concepts from game theory. More details can be found in standard textbooks such as \cite{OsbRub94,weibull1997evolutionary}.

Formally, a (noncooperative) game involves a set of {\em players} $\mathcal{P}=\{1,\dots,n\}$, each associated with a set of {\em pure strategies} ${S}_i=\{1,\dots,m_i\}$ ($i=1 \ldots n)$. A vector of pure strategies $s=(s_1, \ldots , s_n)^T$ is called a {\em strategy profile} and  the set $S = S_1 \times S_2 \times \ldots \times S_n$ is the {\em pure strategy space} of the game.  The \emph{(combined) payoff function} is a function $\pi:{S}\rightarrow\mathbb{R}^n$ which assigns a real valued payoff $\pi_i(s)\in\mathbb{R}$ to each player $i\in \mathcal{P}$ and (pure) strategy profile $s\in{S}$. This is the payoff that player $i$ gets when players play (pure) strategies $s_1 , \ldots , s_n$.

A \emph{mixed strategy} for player $i\in\mathcal{P}$ is simply a probability distribution over its pure strategy set ${S}_i$. Geometrically, it is a vector $x_i=\left(x_{i1},\dots,x_{im_i}\right)^T$ belonging to the \emph{standard simplex}:
$$
\Delta_i =\left\{x_i\in\mathbb{R}^{m_i}: \displaystyle\sum_{h=1}^{m_i}x_{ih}=1\mbox{, and } x_{ih}\ge 0\mbox{ for all $h \in S_i$}\right\} ~.
$$
Each component $x_{ih}$ denotes the probability with which player $i$ will choose its pure strategy $h$. 
Accordingly, a \emph{mixed strategy profile} $x=\left(x_1,\dots,x_n\right)^T$ is defined as a vector of mixed strategies, each \mbox{$x_i\in\Delta_i$} representing the mixed strategy assigned to player \mbox{$i\in\mathcal{P}$}. Each mixed strategy profile lives in the \emph{mixed strategy space} defined as 
$
\Theta=\Delta_1 \times \Delta_2 \times \ldots \times \Delta_n  ~.
$

In non-cooperative game theory we make the assumption that players choose their strategies in total ignorance of the other players' choices. Under this independence assumption, the probability that a given pure strategy profile $s \in S$ will be used when mixed strategy profile $x \in \Theta$ is played becomes:
\begin{equation}
	x(s)=\prod_{i=1}^n x_{is_i}
\end{equation} 
and hence the {\em expected  payoff} for player~$i$ associated to $x \in \Theta$ can be written as: 
\begin{equation}
	u_i(x)=\sum_{s \in S} x(s) \pi_i(s) .
\end{equation} 

Following standard game-theoretic notations, let $(x_i,y_{-i})\in\Theta$ denote the strategy profile where player $i$ plays strategy \mbox{$x_i\in\Delta_i$} whereas the other players $j\in\mathcal{P}\setminus\{i\}$ play based on the strategy profile $y\in\Theta$. 
A mixed strategy profile $x^*=(x_1^*,\dots,x_n^*) \in \Theta$ is said to be a {\em Nash equilibrium} if no player has an incentive to unilaterally deviate from it or, more formally, if
\begin{equation} 
	u_i(x_i^*,x_{-i}^*)\ge u_i(x_i,x_{-i}^*)
\label{eq:Nash}
\end{equation}
for all $i\in\mathcal{P}$ and $x_i\in S_i$. If the inequalities hold strictly, $x^*$ is called a {\em strict} Nash equilibrium. Nash equilibrium is a key concept in game theory. In his celebrated paper \cite{nash1951non}, John Nash proved that any non-cooperative game has at least one (mixed) Nash equilibrium.

\vspace{-0.2cm}
\subsection{Jigsaw puzzle games}
\vspace{-0.2cm}

We now cast the puzzle-solving problem as a non-cooperative multiplayer game.

According to our formulation, the set of players $\cal P$   coincides with the set of all puzzle pieces, and the set of pure strategies associated to each player is the set of all possible triplets $s_i = (x_i,y_i,\theta_i)$, where $(x_i,y_i)$ refers to planar coordinates and $\theta_i$ to a rotation angle.
Of course, to reduce the computational burden, in our experiments we discretize the strategy space and make a number of simplifying assumptions (see next section for details).

In this paper, we suppose that the payoffs associated to each player are ``additively separable,'' i.e., that for a pure strategy profile $s=(s_1,\dots,s_n)\in S$, the payoff of every player $i\in\mathcal{P}$ can be written as:
\begin{equation}
	\pi_i(s)=\sum_{j=1}^n A_{ij}(s_i,s_j)
\end{equation} where $A_{ij}$ is the \emph{partial payoff} matrix between players $i$ and $j$. 
This makes the proposed game a member of a well-studied subclass of multi-player games, known as \emph{polymatrix games}~\cite{Jan68}.

Here, the payoff function reflects the consensus between pieces given their chosen positions and orientation. 
Accordingly, given two pieces $i,j \in \cal P$ and two pure strategies $s_i=(x_i,y_j,\theta_j)$ and $s_j=(x_j,y_j,\theta_j)$, the payoff $A_{ij}(s_i,s_j)$ measures the compatibility of the following two hypotheses: 
\begin{itemize}
\item $H_1$: ``piece $i$, rotated by $\theta_i$ degrees, is in position $(x_i,y_i)$''
\item $H_2$: ``piece $j$, rotated by $\theta_j$ degrees, is in position $(x_j,y_j$)''
\end{itemize}

Once the jigsaw-puzzle problem is formulated in this way, it is easy to see that Nash equilibria correspond to plausible puzzle solutions. This is indeed motivated by a formal connection established by Miller and Zucker \cite{MilZuc91} between Nash equilibria of polymatrix games and the notion of a consistent labeling in relaxation labeling processes \cite{HumZuc83,Pel97} (see also \cite{khoroshiltseva2021jigsaw}).

\vspace{-0.2cm}
\subsection{Finding Nash equilibria}
\vspace{-0.2cm}

To find a Nash equilibrium of our jigsaw puzzle game we use a well-known class of game dynamics studied in evolutionary game theory \cite{HofSig98,weibull1997evolutionary}.
Under this framework one imagines that the game is played repeatedly, generation after generation, in a population of players and that a selection process acts on the strategy set, resulting in the evolution of the fittest strategies. 

This selection dynamics is commonly modeled by (continuous-time) multi-population {\em replicator dynamics} inspired by Darwin's principle of natural selection. In our polymatrix game setting, this can be written as: 

\begin{equation}
\dot{x}_{ih} = x_{ih}\left(  \pi_{ih}(x) - \sum_{k} x_{ik} \pi_{ik}(x) \right)
\label{eq:repdyncont}
\end{equation}
where $i\in \cal P$ and $h \in S_i$ is a pure strategy associated to player $i$.
Here, 
\begin{equation}
\pi_{ih}(x) = \sum_{j} \sum_{k} x_{jk} A_{ij}(h,k)
\end{equation}
measures the expected payoff of player $i$ when playing strategy $h$ and
$\sum_{k} x_{ik} \pi_{ik}(x)$ is the average population payoff.

The following theorem states that interior limits points of (\ref{eq:repdyncont}) are in one-to-one correspondence to Nash equilibria (see, e.g., \cite{weibull1997evolutionary}) for a proof). This motivates the use of replicator dynamics to find the Nash equilibria of our puzzle games. 

\begin{theorem}{A point $x\in\Theta$ is a limit trajectory of (\ref{eq:repdyncont}) starting from the interior of $\Theta$ if and only if $x$ is a Nash equilibrium. Further, if point $x\in\Theta$ is a strict Nash equilibrium then it is also asymptotically stable.}
\end{theorem}

In our experiments, we used the following discrete-time counterpart of (\ref{eq:repdyncont}),
\begin{equation}
x_{ih}(t+1) = x_{ih}(t)\frac{\pi_{ih}(x(t))}{\sum_{k} x_{ik}(t) \pi_{ik}(x(t))}
\label{eq:repdyndisc}
\end{equation}
which has the same dynamical properties as the continuous one (see~\cite{Pel97} for a detailed analysis).

The computational complexity of finding a Nash equilibrium of a our puzzle game using~(\ref{eq:repdyndisc}) is $\mathcal{O}(kcn^2),$ where $n$ is the number of players (pieces), $c$ is the number of pure strategies (position/orientation) and $k$ is the number of iterations.
In the following section we shall see how to dramatically reduce the computational complexity of the algorithm by using a proper discretization and sparsification strategy.

\vspace{-0.2cm}
\section{Payoff functions}
\vspace{-0.2cm}

\subsection{Partial payoff matrix} 
\vspace{-0.2cm}

\begin{figure}
    \centering
    \includegraphics[width=0.8\textwidth]{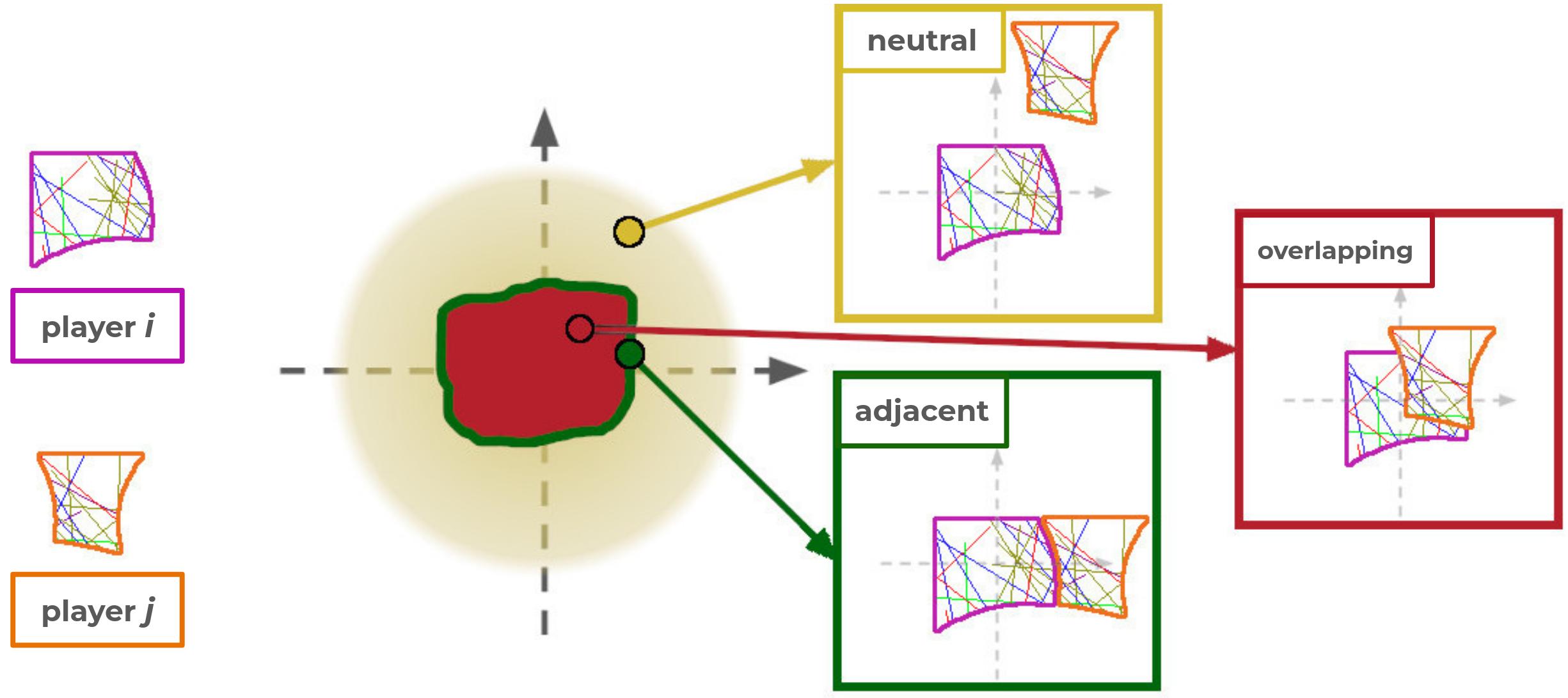}
    \caption{The three regions of the space : neutral (yellow), adjacent (green) and overlapping (red).
    \vspace{-0.1cm}}
    \label{fig:3areas}
\end{figure}

To reduce the strategic space, we define for each pair of players a proximity space $\Gamma$.
In this space the player $i$ assumes the strategy $s_0=(0,0,0)$, while the player $j$ translates and rotates around $i$. The space is divided into three regions: \textit{adjacent}, \textit{overlapping} and \textit{neutral}. The region where players $i$ and $j$ maintain contact without overlapping denotes “legal” positioning, positively contributing to players' gain.
On the contrary, the area in which the players overlap is considered an "illegal" positioning and corresponds to a negative contribution - penalizing the player's payoff.
Other positions, where players are not in direct contact, are labeled "neutral" and do not explicitly affect players' payoffs. 
The concept is illustrated in Fig. \ref{fig:3areas}.

We therefore define the reduced \emph{partial payoff} matrix $\Tilde{A}_{ij}$ which contains the payoffs related to the relative strategies $\gamma \in \Gamma$ of player $j$ against the fixed strategy of player $ i $. The payoff $\Tilde{A}_{ij}(\gamma)$ is formally expressed in Eq. \ref{eq:comp_coef}:

\begin{equation}
    \\\Tilde{A}_{ij}(\gamma) =
    \begin{cases}
       R_{ij\gamma} &  (i \neq j) \land (\gamma \in \text{\textit{adjacent} region})\\
       -1 &  (i = j) \lor  (\gamma \in \text{\textit{overlapping} region})\\
        0 &  (i \neq j) \land (\gamma \in \text{\textit{neutral} region})
    \end{cases}
    \label{eq:comp_coef}
\end{equation}

The compatibility score $R_{ij\gamma}$ evaluates the placement when player $j$ chooses the strategy $\gamma$
and can be calculated based on various factors such as appearance (e.g., colors or line continuity), geometry, or semantic proximity. 
Here we focus on the line-based compatibility.

\vspace{-0.2cm}
\subsection{Line-based Compatibility Measure}
\label{sec:compatibility}
\vspace{-0.2cm}

The compatibility score $R_{ij\gamma}$ reflects the quality of the match based on the characteristics of the two players, approaching $1$ for the best strategy and approaching $0$ for strategies that do not significantly benefit the players.
We decide to use the lines as characteristics of each player in the game and calculate the compatibility score $R_{ij\gamma}$ based on lines only.

The score is calculated according to the Gestalt principle of good continuation \cite{Kof63} along the lines of both players.
The compatibility score between two players $i$ and $j$ for a relative strategy $\gamma$ is high if the lines of $j$ are continuations of the lines of $i$ and vice versa. A line $r$ of a player $i$ denoted as $l_{i,r}$ is represented by a start and end point, an angle $\alpha_{r}$ and a semantic category $c_r$.

First, we estimate the cost as a measure of the effort needed to continue a pair of lines, which we denote as $C_L$ and describe in Eq. \ref{eq:line_cost}. 
This cost is zero when the lines perfectly continue each other and reaches high values when the lines are not aligned. 

\vspace{-0.3cm}
\begin{equation}
    C_L(l_{i,r}, l_{j,s}) = 
    \begin{cases}
       {dist} (l_{i,r}, l_{j,s}) & \textnormal{if } |\alpha_{r}-\alpha_{s}| < \epsilon_{a} \land c_r = c_s \\
       cost_{max} & otherwise
    \end{cases}
    \label{eq:line_cost}
\end{equation}

The $dist(l_{i,r}, l_{j,s})$ is the smallest distance between any of the two end points of each line, $cost_{max}$ is a constant value related to the puzzle, $\epsilon_{a}$ is the tolerance of the angle between two lines.
 
The total cost of matching the set of lines of player $i$ (denoted as $L_i$) and those of player $j$ (denoted as $L_j$) is estimated as  described in Eq. \ref{eq:c_tot}. The first component $C_{\mathbf{LAP}}$ is obtained by solving the associated Linear Assignment Problem (\texttt{LAP})\cite{brualdi2006combinatorial} and the second $\phi_{ij}$ is a penalty term for the unmatched lines.

\begin{equation}
    C_{tot}(L_{i}, L_{j} | \gamma) = C_{\mathbf{LAP}} (L_{i}, L_{j} | \gamma) + \phi_{ij}
    \label{eq:c_tot}
\end{equation}

The cost of the solution \texttt{LAP} consists of the sum of the costs of matching pairs of lines selected in such a way that the cost of the assignment is minimized.
\begin{equation}
    C_{\mathbf{LAP}} (L_{i}, L_{j} | \gamma) = \sum_{r\in(1,|L_i|)} C_L(l_{i,r}, l_{j,s(r)} | \gamma)
    \label{eq:c_lap}
\end{equation}

\begin{figure}
    \centering
    \includegraphics[width=\textwidth]{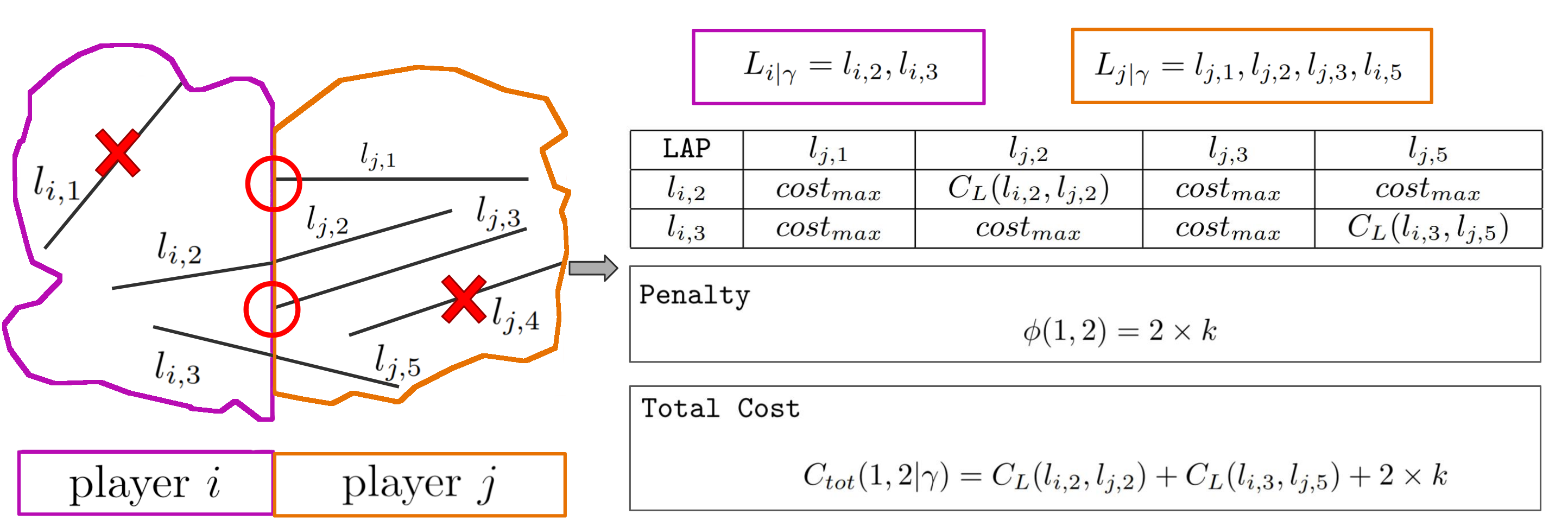}
    \caption{An example of the computation of the cost for the players $i$ and $j$ given a candidate $\gamma$ strategy. 
    The two sets $L_{i|\gamma}$ and $L_{j|\gamma}$ are created by selecting only lines which intersects the opposite player, and the total cost consists of the sum of the cost of the corresponding \texttt{LAP} and the penalty for unmatched lines. 
    \vspace{-0.1cm}
    } 
    \label{fig:cost_comp}
\end{figure}

We denote the assignment solution as $s(r)$, which associates the $s(r)$-line in $L_j$ to the $r$-line in $L_i$. 
We use the $C(\cdot | \gamma)$ notation to dynamically create the sets of lines such as $L_{i | \gamma}$ by selecting only the subset of lines of $L_{i}$ which are coherent to the current relative strategy $\gamma$ (lines $l_ {i,1}$ and $l_ {j,4}$ of example in Fig. \ref{fig:cost_comp}).
The \texttt{LAP} solution assigns a number of matches based on the cardinality of the smallest set. In the presence of unassigned lines we penalize the solution by adding the penalty term $\phi=u_{l} \cdot k$, where $k$ is the constant factor and $u_l$ is the number of unassigned lines

\vspace{-0.7cm}
\vspace{-0.3cm}

The compatibility score is a measure of the quality of the continuation of all lines of both players given the relative strategy $\gamma$. 
Therefore, it will be inversely proportional to the total cost of matching the two sets, as described in Eq. \ref{eq:cij}:

\begin{equation}
R_{ij\gamma} = 1 - \frac{\mathnormal{min}(C_{tot}(i,j,\gamma), \tau)}{\tau}
\label{eq:cij}
\end{equation}

The cost is truncated and normalized using $\tau$ which is based on the best compatibility scores obtained by all players.
In doing so, we discard the strategies that obtained lower scores, sparsifying the partial payoff matrix $\Tilde{A}_{ij}(\gamma)$ and reducing the computational complexity.

\vspace{-0.2cm}
\section{Experiments}
\vspace{-0.2cm}

To demonstrate the effectiveness and flexibility of our method, we conducted an extensive evaluation on different types of puzzles. 
Concretely, we evaluate our approach on:

\begin{enumerate}
\vspace{-0.2cm}

    \item Puzzles with \emph{squared pieces with known piece orientation} 
    \item Puzzles with \emph{polygonal shapes with unknown piece orientation} 
    \item Puzzles with \emph{irregular shapes with unknown piece orientation} 
\end{enumerate}

\vspace{-0.2cm}
For all categories, we experimented with synthetic data generated by randomly drawing lines on blank images and real data in the form of vector-based geographical maps.

\subsection{Datasets}
\vspace{-0.2cm}
The creation of a large scale dataset for puzzle solving is a challenging task and there is not a standard dataset with different types of pieces for the puzzle solving community. 
We decided to use two sets of puzzles to evaluate our approach: synthetically generated puzzles and puzzles obtained cutting real-world data into pieces.
\vspace{-0.6cm}

\subsubsection{Synthetic Lines Generation}

We developed a puzzle generator pipeline to create a dataset with fine-grained control on different scenarios, including the number of pieces, their shapes (squared, polygonal, irregular), and the characteristics of lines (number and types of lines) in an image. 

For the puzzles used in this work, we created images with 50 random lines, each assigned a semantic category. 
In the baseline scenario of squared pieces, we created 30 images using a single semantic category.
In the polygonal and irregular scenarios, we created 10 images using $5$ different categories rendered with distinct colors on the associated images.

\vspace{-0.6cm}

\subsubsection{Puzzles from Real Data}

Since we tackle the novel task of \emph{line-based} puzzle solving, we created a dataset that consists of vector maps. 
Vector maps are a real-world example of line-based real-world data that exhibit a wide variety of line types and shapes.
We gathered vector maps from the Openstreetmaps\footnote{OpenStreetMap: \url{https://www.openstreetmap.org}\label{osm}} to create a test set of $10$ images.
We processed them to extract coordinates, angles and the semantic category (related to the mapped object) 
and converted them to raster images to create the puzzle images.

\begin{figure}
    \centering
    \begin{subfigure}{0.32\textwidth}
        \centering
        \includegraphics[width=0.7\textwidth]{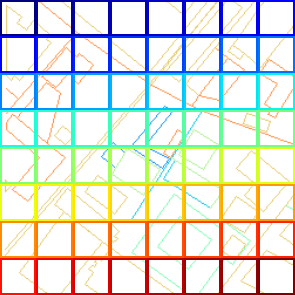}
        \caption{Puzzle with square pieces from Honolulu}
        \label{fig:first}
    \end{subfigure}
    \hfill
    \begin{subfigure}{0.32\textwidth}
        \centering
        \includegraphics[width=0.7\textwidth]{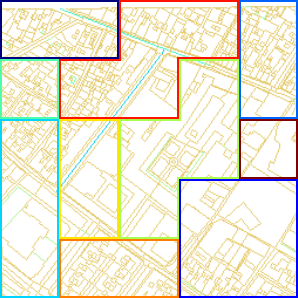}
        \caption{Puzzle with polygonal pieces from Barcelona}
        \label{fig:second}
    \end{subfigure}
    \hfill
    \begin{subfigure}{0.32\textwidth}
        \centering
        \includegraphics[width=0.7\textwidth]{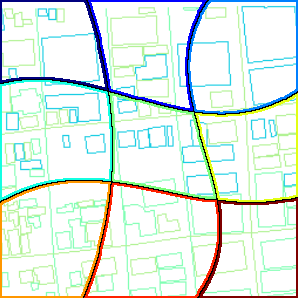}
        \caption{Puzzle with irregular pieces from New York.}
        \label{fig:third}
    \end{subfigure}
    \caption{Some examples of puzzles using  real vector maps from Openstreetmaps{\color{red}\textsuperscript{1}}.
    \vspace{-0.2cm}
    }
    \label{fig:puzzle_types_maps}
\end{figure}
\vspace{-0.75cm}

\subsubsection{Pieces Creation}
The puzzles aim to be a benchmark for the evaluation of different types of algorithms, therefore we created $3$ types of puzzle pieces. 
We created squared pieces by evenly dividing the image space, polygonal pieces using handcrafted patterns (for a finer control on the solution) and irregular pieces by partitioning the image using horizontal and vertical curves. 
Example of puzzles created from real data using these techniques are visible in Fig. \ref{fig:puzzle_types_maps}.

\vspace{-0.2cm}
\subsection{Evaluation metrics}
\vspace{-0.2cm}

In the squared pieces environment, we use known metrics \cite{cho2010probabilistic, pomeranz2011fully}, namely the direct (\textbf{D}) metric, which compute the ratio between the number of pieces correctly assembled and the total number of pieces in the puzzle, and the neighbors (\textbf{N}) metric, which compute the ratio of correct relative neighbors (independently of their absolute position).
For the irregular puzzles, we report only the direct (\textbf{D}) metric modified to accept the correct assembly with a tolerance threshold.

\vspace{-0.2cm}
\subsection{Results} 
\vspace{-0.2cm}

The performance obtained in the different datasets demonstrates the generalization capability and the accuracy of our method in the puzzle solving task. 
\vspace{-0.3cm}
\begin{table}
    \centering
    \begin{tabular}{l@{\hspace{2em}}cc|cc}  \toprule
    \multicolumn{1}{c}{} & \multicolumn{4}{c}{\textbf{Squared Pieces}}\\\cmidrule(lr){2-5}

    \multirow{2}{*}{\textbf{Algorithm}} & \multicolumn{2}{c}{\textbf{Synthetic}} & \multicolumn{2}{c}{\textbf{Maps}} \\\cmidrule(lr){2-3} \cmidrule(lr){4-5}
    & \textbf{D ($\uparrow$)}& \textbf{N ($\uparrow$)}& \textbf{D ($\uparrow$)}&  \textbf{N ($\uparrow$)}\\ \hline 
     Ours & \textbf{0.99} & \textbf{0.93} & \textbf{0.203} & \textbf{0.268} \\
     Pomeranz et al. \cite{pomeranz2011fully} & 0.014
 & 0.241 & 0.020 & 0.158 \\ 
     Paikin et al. \cite{paikin2015solving} & 0.022 & 0.197 & 0.060 & 0.208\\ 
     Positional Diffusion
     \cite{giuliari2023positional} & 0.072 & - & 0.035 & - \\
    \bottomrule
    \end{tabular}
    \vspace{0.1cm}
     \caption{Quantitative results on puzzles with squared pieces. }    
    \label{tab:squared_results}
\end{table}
\vspace{-1cm}

We compare our approach with Pomeranz et al. \cite{pomeranz2011fully}, Paikin et al. \cite{paikin2015solving} and Positional Diffusion\footnote{We use their implementation (available at the time) for the results, which does not provide values for the Neighbors metric.} \cite{giuliari2023positional} for squared pieces with known orientation in Table \ref{tab:squared_results} and with Derech et al. \cite{DERECH2021108065} for arbitrarily shaped pieces (polygonal and irregular) with unknown orientation in Table \ref{tab:irregular_results}.
To ensure a fair comparison, we need to take into consideration some of the differences between the algorithms, since previous works were designed to tackle colored images. As a general rule, we render our lines with different colors based on the semantic category to enable a possible color match. 
Moreover, in the squared pieces scenario, we feed the input image (therefore fixing the final shape of the puzzle) to both competitor approaches, while in our case we just fix one anchor and solve the puzzle around it.
In the irregular scenario, the method proposed by Derech et al. requires an extrapolated version of each piece: we built a \textit{perfect} extrapolated version for all pieces in all puzzles to allow a correct match between overlapping parts. 
Even with these settings, our approach is the only one able to achieve a satisfactory solution on the different puzzles.

\subsubsection{Squared Pieces}

The performance obtained for puzzles with square pieces are presented in Table \ref{tab:squared_results}. Our approach significantly outperforms competitors in terms of all metrics, establishing a new baseline for this novel task.

\begin{table}[!h]
    \centering
    \begin{tabular}{c c c c c}
    \toprule
    Groundtruth & Ours & Pomeranz \cite{pomeranz2011fully} & Paikin\cite{paikin2015solving} & PD \cite{giuliari2023positional, scarpellini2024diffassemble} \\ \midrule
    \includegraphics[width=0.15\textwidth]{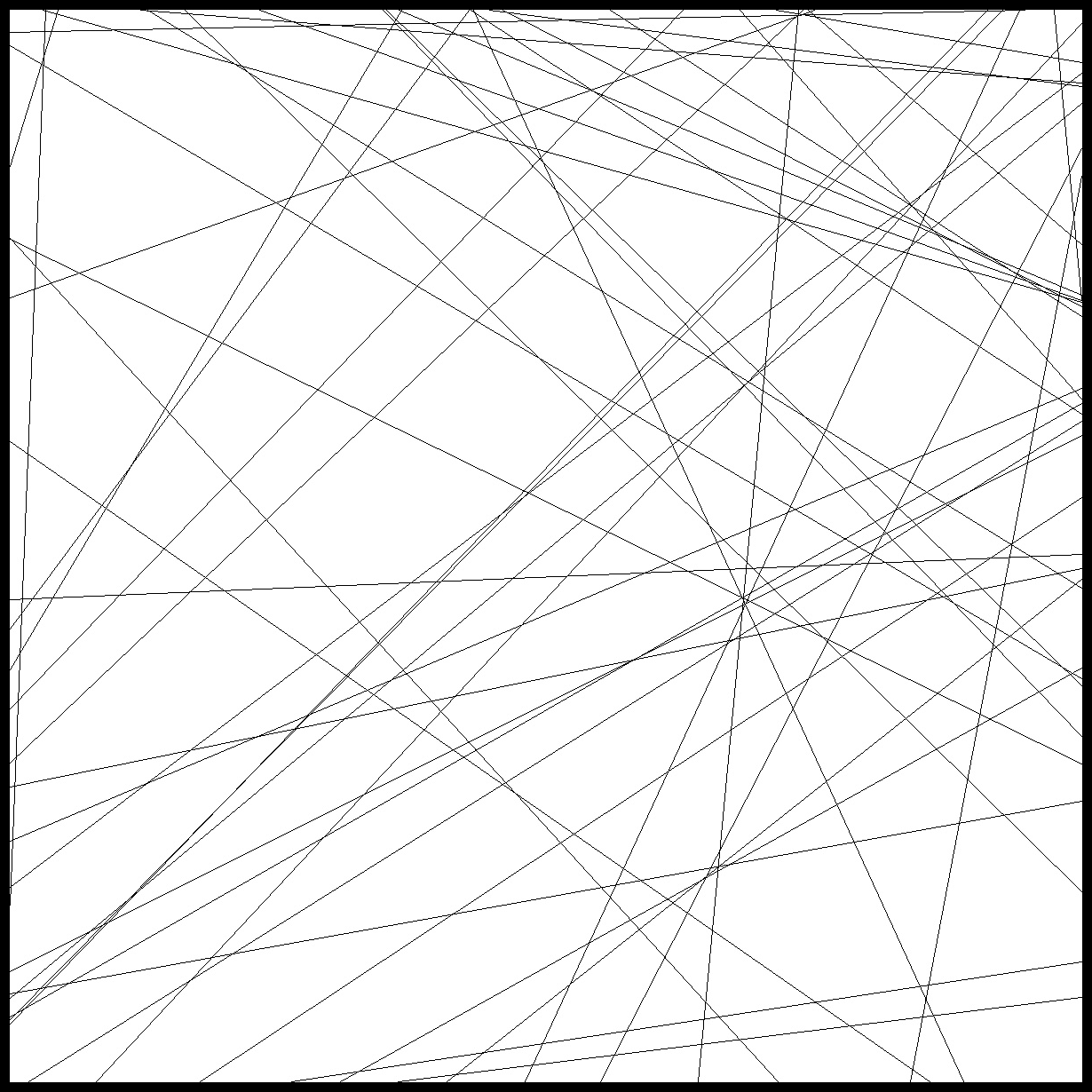} & 
    \includegraphics[width=0.15\textwidth]{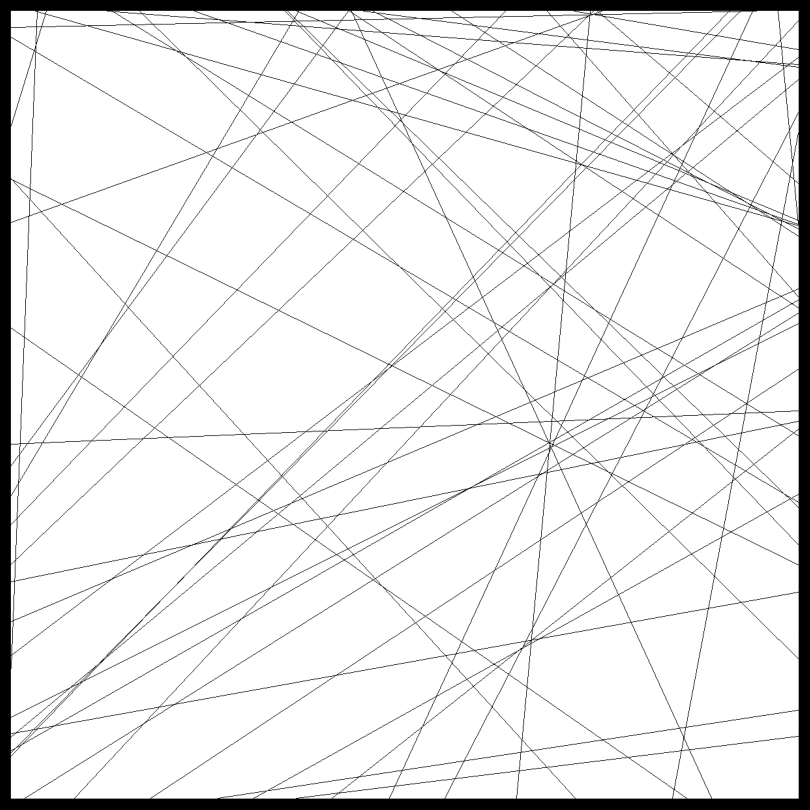} & 
    \includegraphics[width=0.15\textwidth]{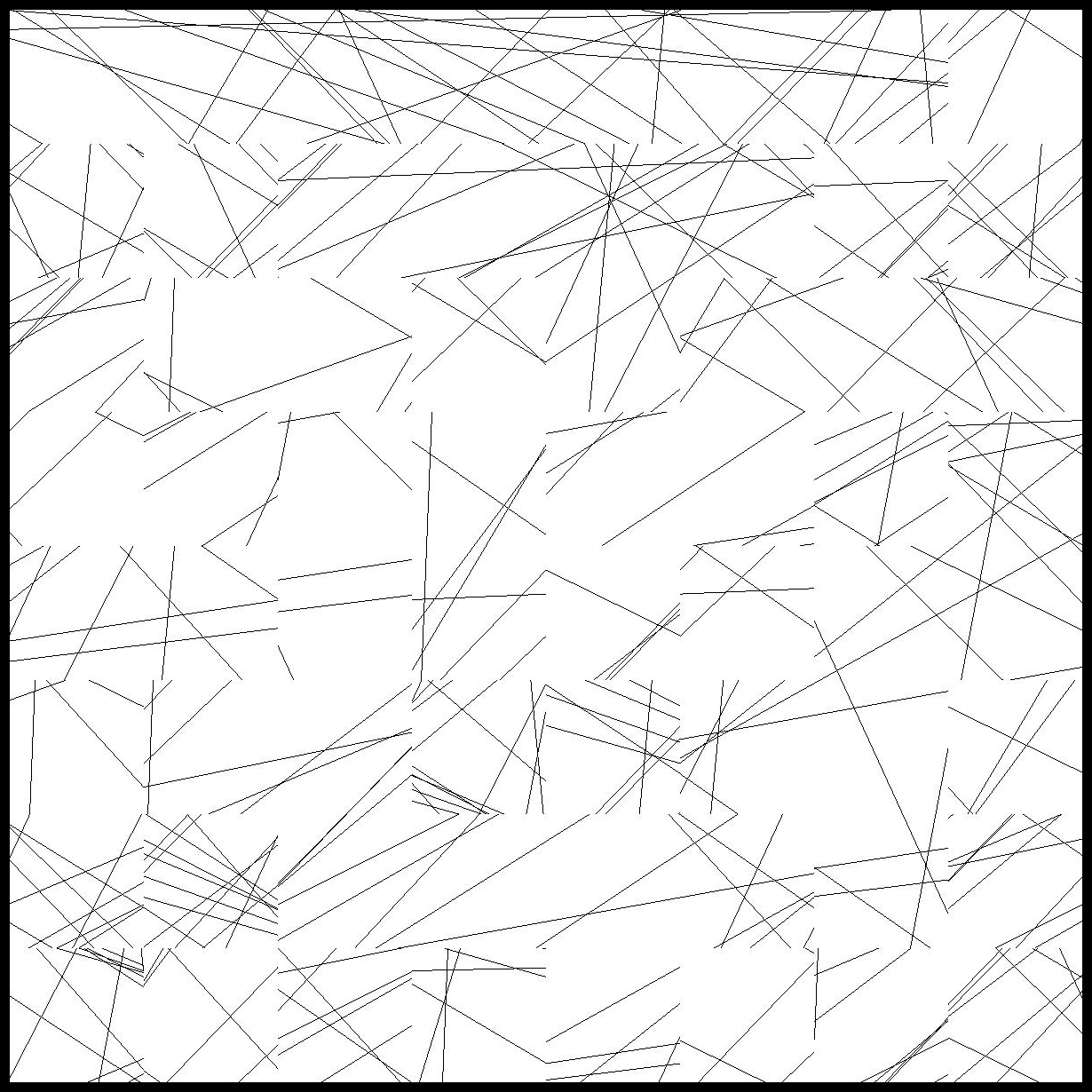} & \includegraphics[width=0.15\textwidth]{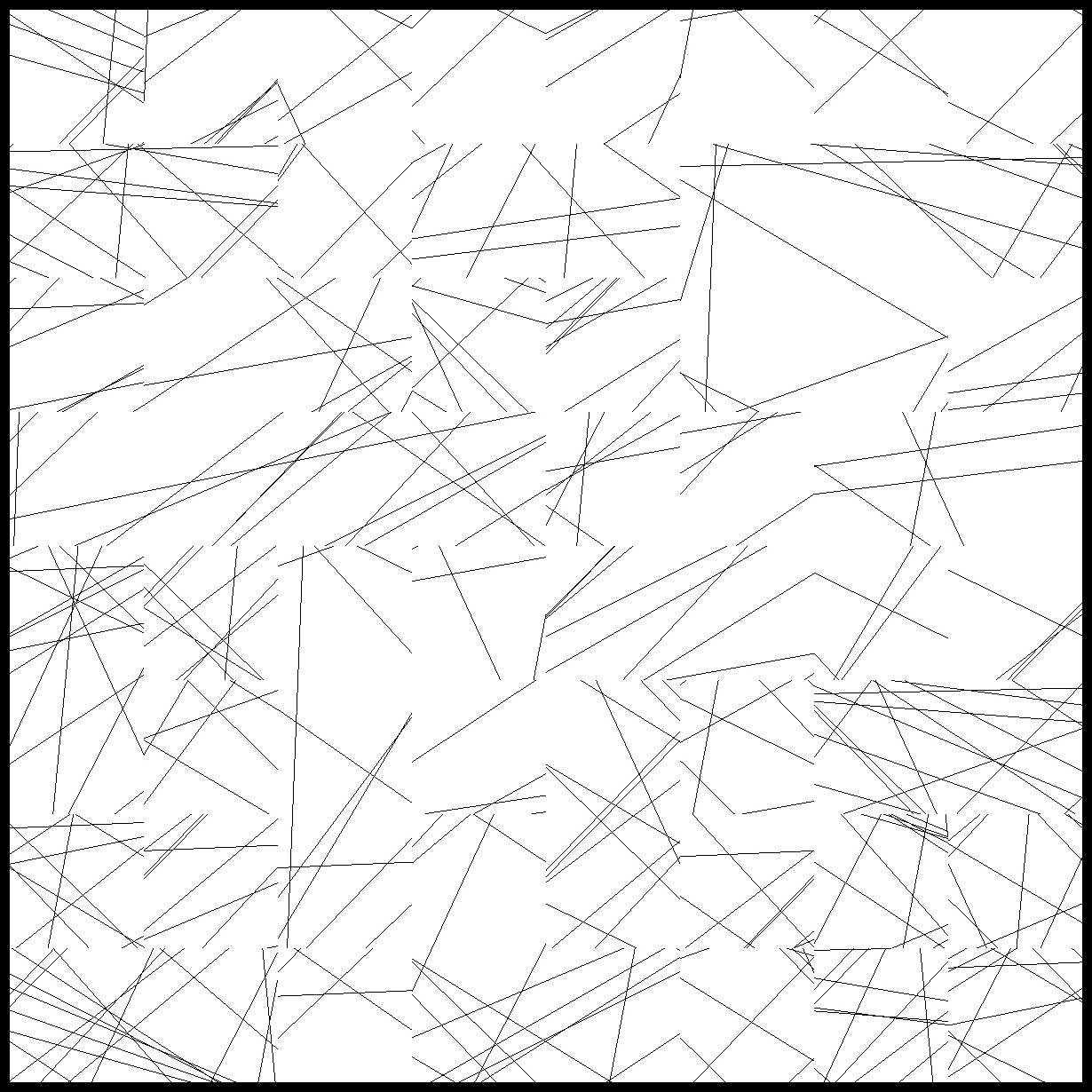} &
    \includegraphics[width=0.15\textwidth]{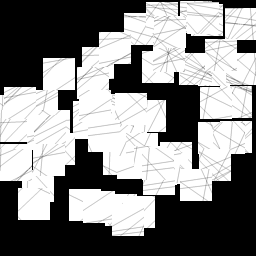} \\    
    \includegraphics[width=0.15\textwidth]{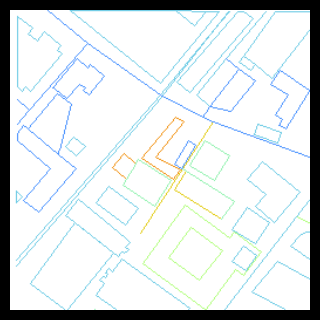} & \includegraphics[width=0.15\textwidth]{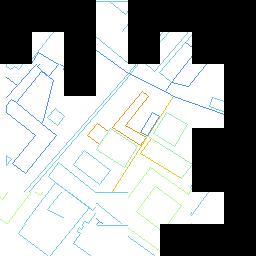} &
    \includegraphics[width=0.15\textwidth]{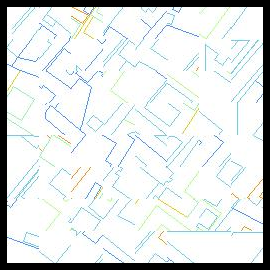} & 
    \includegraphics[width=0.15\textwidth]{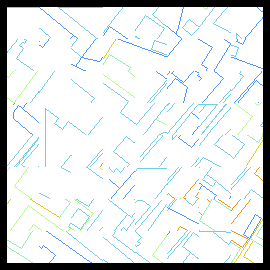} & \includegraphics[width=0.15\textwidth]{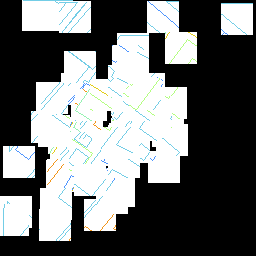} \\
    \includegraphics[width=0.15\textwidth]{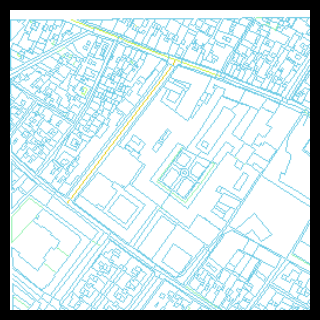} & \includegraphics[width=0.15\textwidth]{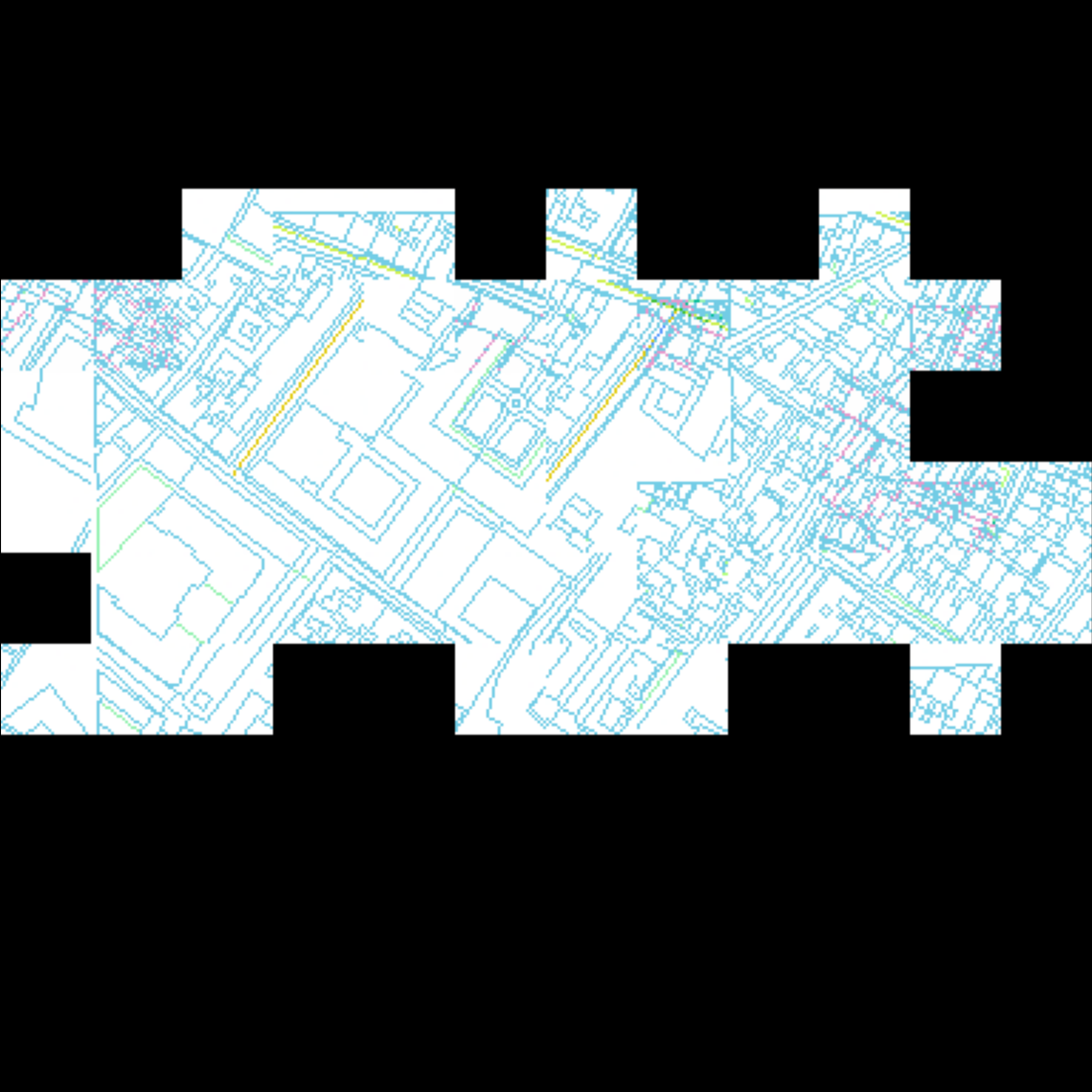} &
    \includegraphics[width=0.15\textwidth]{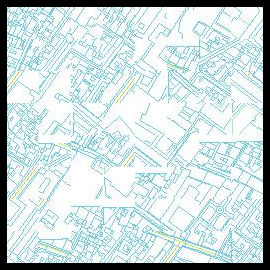} & 
    \includegraphics[width=0.15\textwidth]{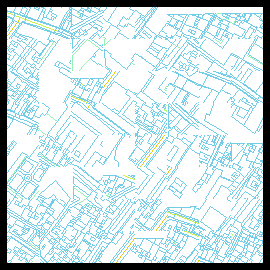} & \includegraphics[width=0.15\textwidth]{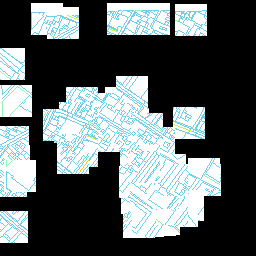} \\
    \bottomrule
    
    \end{tabular}
    \caption{Qualitative Results on squared pieces. 
    \vspace{-0.5cm}
    }
    \label{tab:squared_results_table}
\end{table}

The puzzles from city maps images were created as a novel challenging benchmark, alternating textureless regions with small area with an large density of lines and creating a challenge for any kind of reassembly algorithm.

The accuracy of our approach drops significantly in these puzzles, primarily due to the effect of early wrong choices during reconstruction. In fact, we see success cases where the algorithm reconstructs a large part of the map (as in the second row of Table \ref{tab:squared_results}) and failure cases where the algorithm reconstructs only a smaller portion of the map (and not in the correct relative position). 
None of the state-of-the-art algorithms were able to solve these puzzles.
\vspace{-0.3cm}

\begin{table}[ht]
    \centering
    \begin{tabular}{l@{\hspace{1.6em}}ccccc@{\hspace{1em}}cccc}  \toprule
    & \multicolumn{4}{c}{\textbf{Polygonal Pieces}} & & \multicolumn{4}{c}{\textbf{Irregular Pieces}} \\\cmidrule(lr){2-5} \cmidrule(lr){7-10}
    \multirow{2}{*}{\textbf{Algorithm}} & \multicolumn{2}{c}{\textbf{Synthetic}} & \multicolumn{2}{c}{\textbf{Maps}} & & \multicolumn{2}{c}{\textbf{Synthetic}} & \multicolumn{2}{c}{\textbf{Maps}} \\\cmidrule(lr){2-5} \cmidrule(lr){7-10}
    & \multicolumn{2}{c}{\textbf{D ($\uparrow$)}} & \multicolumn{2}{c}{\textbf{D ($\uparrow$)}} & & \multicolumn{2}{c}{\textbf{D ($\uparrow$)}} & \multicolumn{2}{c}{\textbf{D ($\uparrow$)}} \\\midrule
    Ours ({$90$\textdegree~rotations}) & \multicolumn{2}{c}{\textbf{0.99}} & \multicolumn{2}{c}{\textbf{0.51}} & &  \multicolumn{2}{c}{\textbf{0.622}} & \multicolumn{2}{c}{\textbf{0.283}} \\ 
    Derech et al. \cite{DERECH2021108065} & \multicolumn{2}{c}{0.25} & \multicolumn{2}{c}{0.21} & &  \multicolumn{2}{c}{0.11} & \multicolumn{2}{c}{0.11} \\\bottomrule
    \end{tabular}
    \vspace{0.1cm}
    
    \caption{Quantitative results on polygonal and irregular pieces.
    \vspace{-0.5cm}
    }
    \label{tab:irregular_results}
\end{table}

\vspace{-0.6cm}
\subsubsection{Polygonal and Irregular Pieces}

Our approach tackles without modification any kind of puzzle, and we report the performance on irregular pieces in Table \ref{tab:irregular_results}. 
Again, the accuracy hints that our method is a step in the right direction while the method developed from Derech et al. \cite{DERECH2021108065} struggle to reconstruct the puzzle in both the polygonal and the irregular case.

Our accuracy is significantly higher in the polygonal case, due to the exact definition of the grid. 
Due to the handcrafted nature of the polygonal pieces, we can guarantee that the correct transformation lies within the grid used to define the strategies of our players. 
This is exploited by our algorithm which obtains an almost perfect score on the synthetic polygonal case and correctly assembles more than half of the polygonal pieces, scoring even better in the case of squared pieces on the same maps.
Although the polygonal pieces offer strong shape clues, they seem not to be sufficient for the method by Derech et al. \cite{DERECH2021108065} which fails to assemble them.
\vspace{-0.5cm}

\begin{table}[h]
    \centering
    \begin{tabular}{ c c c | c c c}
    & \multicolumn{4}{c}{\textbf{Polygonal Pieces}} & \\
    \midrule
    Groundtruth & Ours & Derech et al. \cite{DERECH2021108065} & Groundtruth & Ours & Derech et al. \cite{DERECH2021108065} \\
    \midrule
    \includegraphics[width=0.16\textwidth]{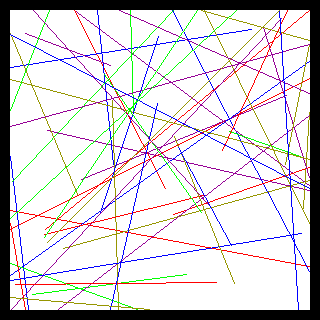} & \includegraphics[width=0.16\textwidth]{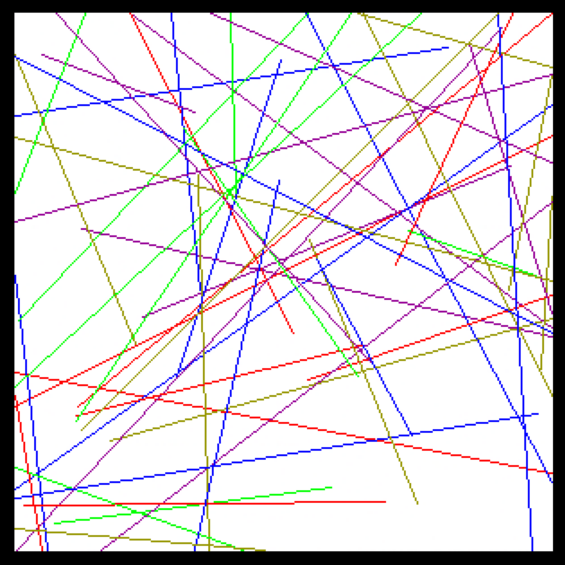} 
    & \includegraphics[width=0.16\textwidth]{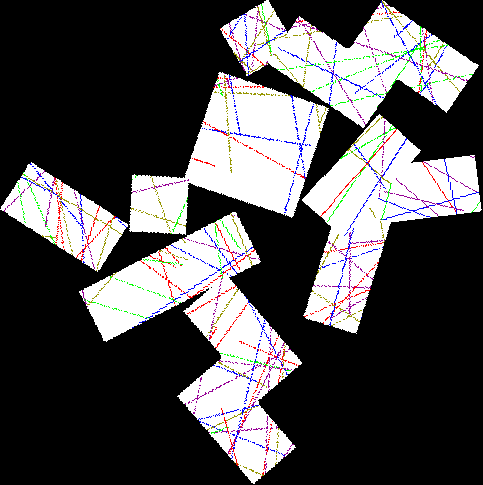} &
    \includegraphics[width=0.16\textwidth]{imgs/qualit_eval/real/maps_image_00000_b.png} &
    \includegraphics[width=0.16\textwidth]{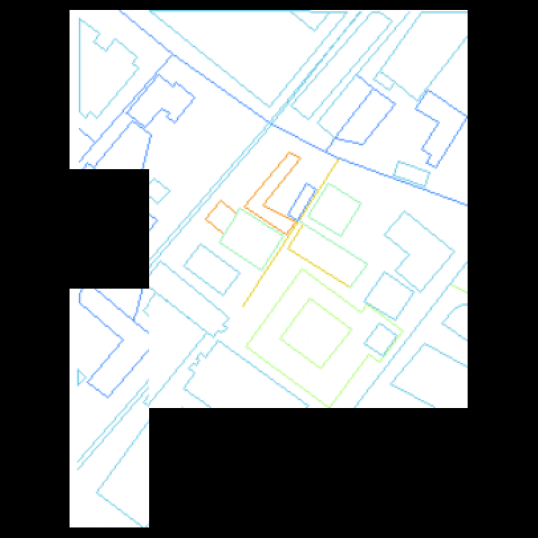} & 
    \includegraphics[width=0.16\textwidth]{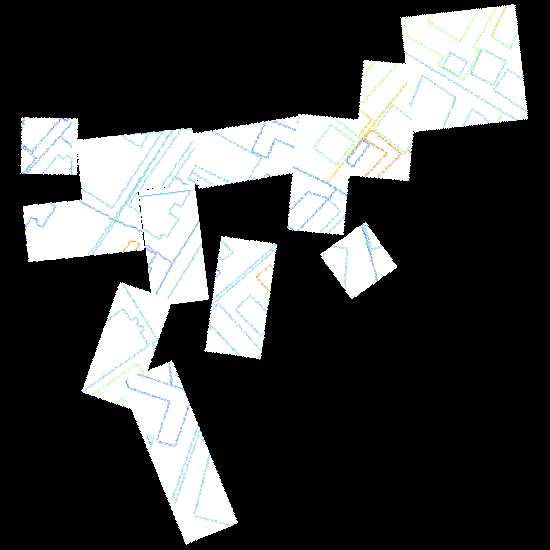} \\
    \includegraphics[width=0.16\textwidth]{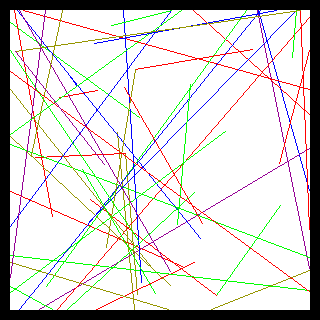} & \includegraphics[width=0.16\textwidth]{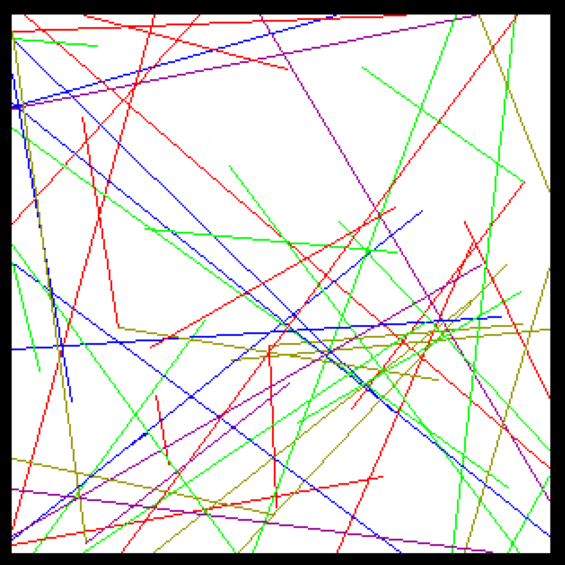} &
    \includegraphics[width=0.16\textwidth]{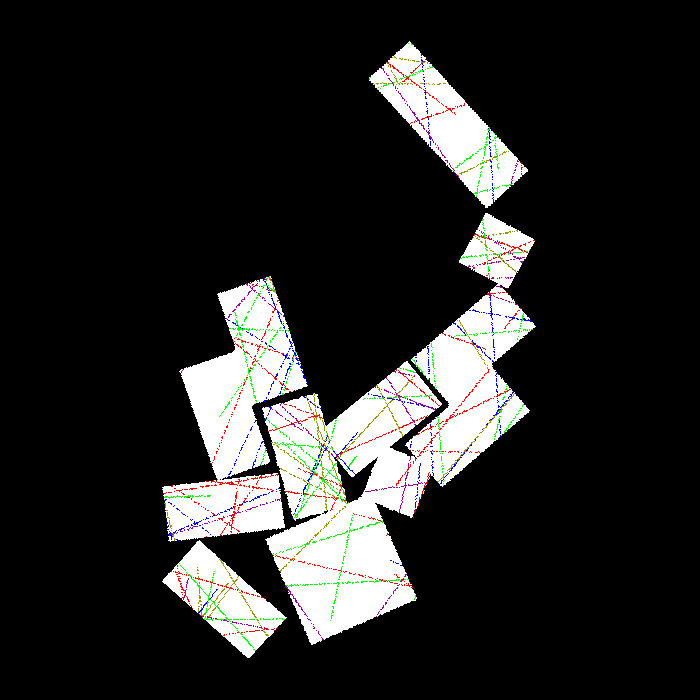} &
    \includegraphics[width=0.16\textwidth]{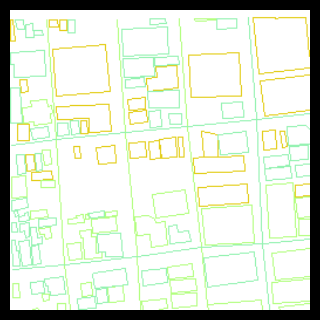} & 
    \includegraphics[width=0.16\textwidth]{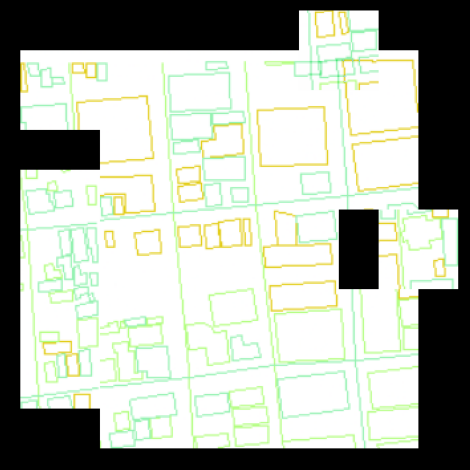} & \includegraphics[width=0.16\textwidth]{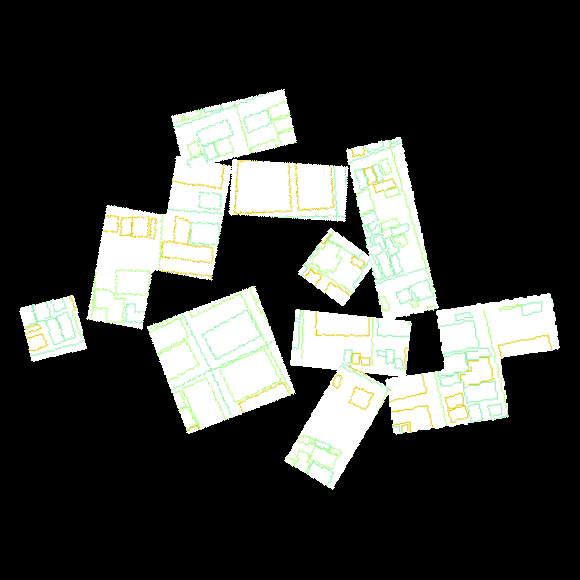} \\ \bottomrule 
    & \multicolumn{4}{c}{\textbf{Irregular Pieces}} & \\
    \midrule
    Groundtruth & Ours & Derech et al. \cite{DERECH2021108065} & Groundtruth & Ours & Derech et al. \cite{DERECH2021108065} \\
    \midrule
    \includegraphics[width=0.16\textwidth]{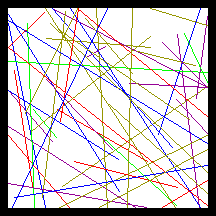} & 
    \includegraphics[width=0.16\textwidth]{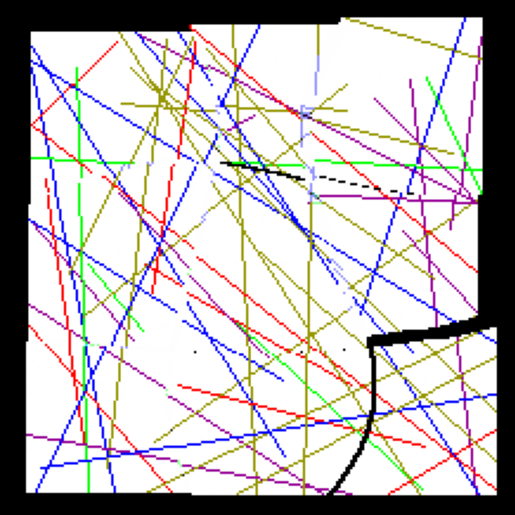} & 
    \includegraphics[width=0.16\textwidth]{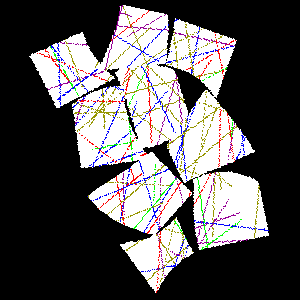} &
    \includegraphics[width=0.16\textwidth]{imgs/qualit_eval/real/maps_image_00000_b.png} & 
    \includegraphics[width=0.16\textwidth]{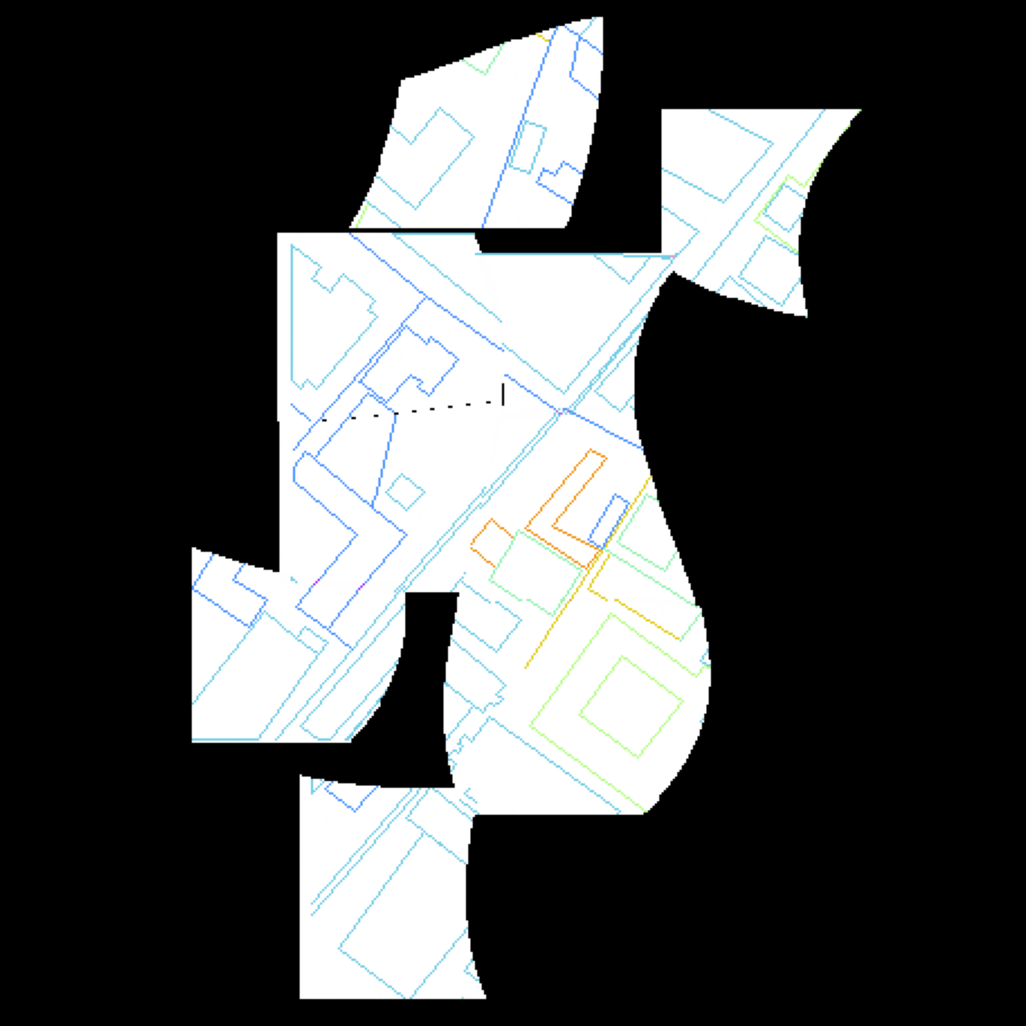} & 
    \includegraphics[width=0.16\textwidth]{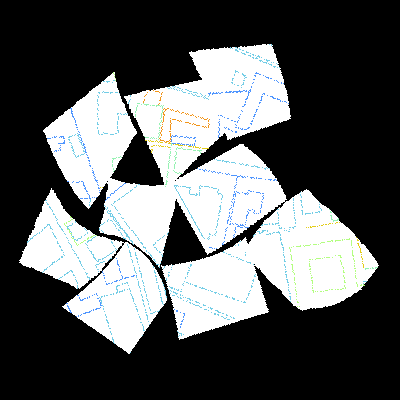} \\ 
    \includegraphics[width=0.16\textwidth]{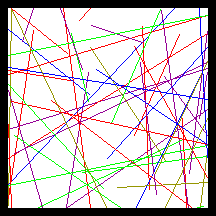} & 
    \includegraphics[width=0.16\textwidth]{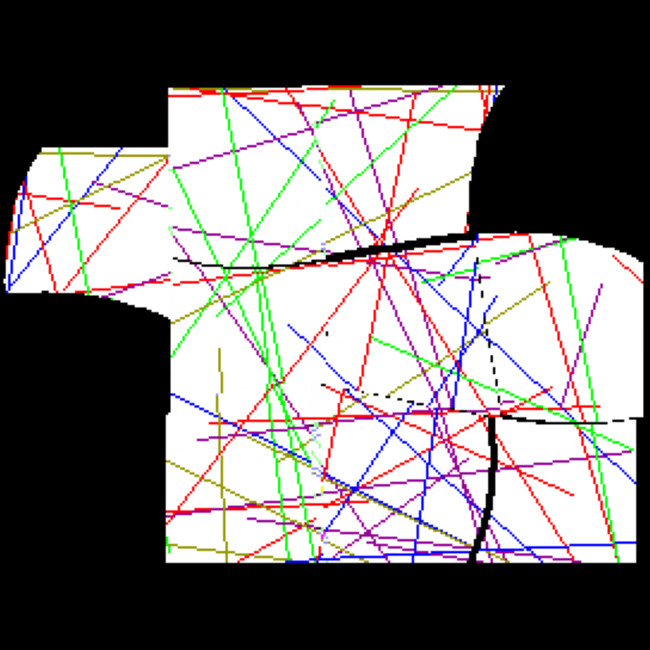} & 
    \includegraphics[width=0.16\textwidth]{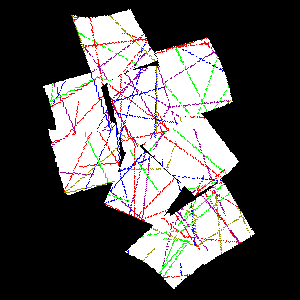} &
    \includegraphics[width=0.16\textwidth]{imgs/qualit_eval/real/maps_image_00002_b.png} & 
    \includegraphics[width=0.16\textwidth]{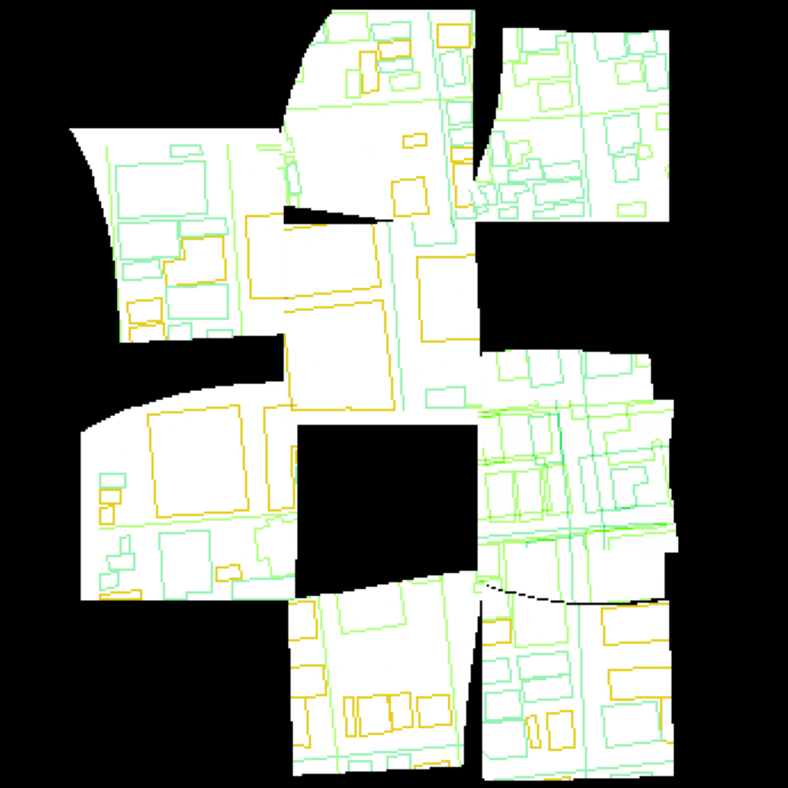} & 
    \includegraphics[width=0.16\textwidth]{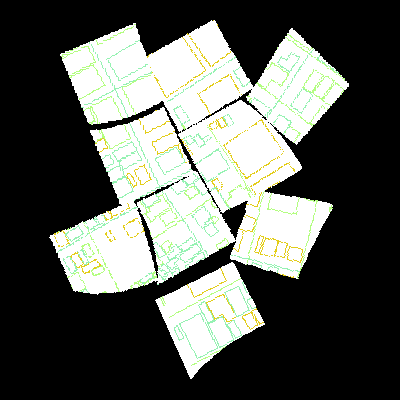} \\ 
    \bottomrule
    \end{tabular}
    \caption{Qualitative results on the polygonal and irregular pieces. Reconstructed puzzles are cropped for a better visualization.
    \vspace{-0.5cm}}
    \label{tab:polygonal_qual}
\end{table}
\vspace{-0.3cm}

In the irregular case, our accuracy is lower, pointing out that the players did not choose the best strategy, or the pieces were not in the absolute correct position.
Although our method achieves sub-optimal accuracy in this scenario, we see that the alternative solutions found are plausible. 

Visual inspection of the reconstructions gives us further insights: our approach finds plausible solutions \textit{which satisfy the principle of good continuation of the lines} without solving the puzzle in its original squared form, resulting in \emph{wrong} alignment \emph{with respect to the metric}.
We want to strengthen here that our algorithm does not know the final shape of the puzzle (which could be helpful, for example in the second synthetic puzzle shown here, to assemble the last piece), as it was designed for a real-world scenario, where the shape of the final puzzle is unknown.

\begin{figure}[h]
    \centering
    \begin{subfigure}{0.411\textwidth}
        \includegraphics[width=\textwidth]{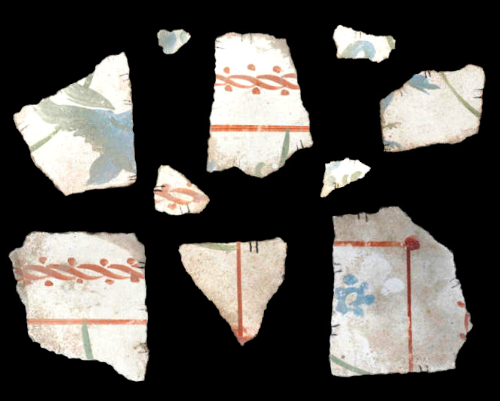}
        \caption{Example of irregular fragments.}
        \label{fig:second}
    \end{subfigure}
    \begin{subfigure}{0.33\textwidth}
        \includegraphics[width=\textwidth]{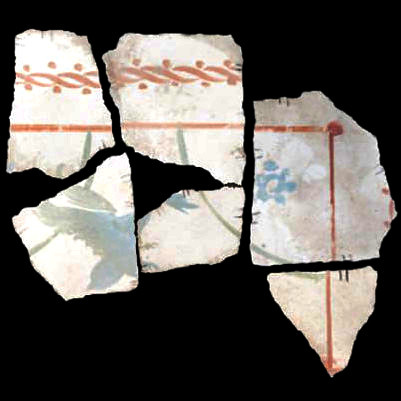}
        \caption{Our solution.}
        \label{fig:our_repair}
    \end{subfigure}
    \begin{subfigure}{0.239\textwidth}
        \includegraphics[width=\textwidth]{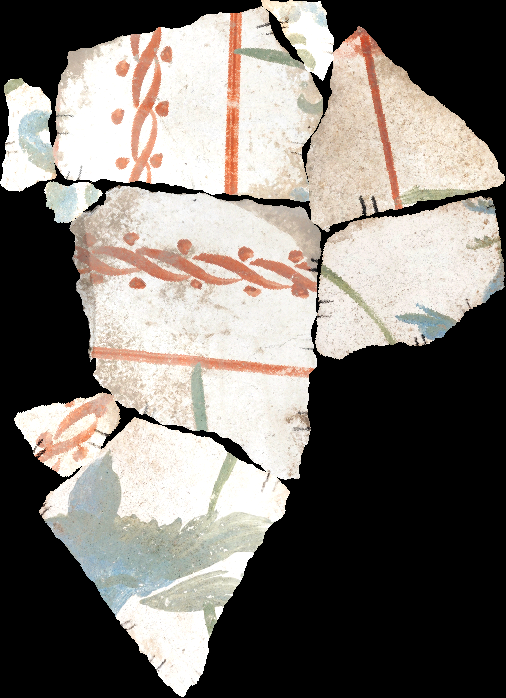}
        \caption{Solution from \cite{DERECH2021108065}.}
        \label{fig:first}
    \end{subfigure}
    \caption{An example of a real archaeological puzzle composed of irregular fragments.}
    \label{fig:repair_puzzle}
\end{figure}

Finally, we provide our results on a case study consisting of a (subset of a) puzzle composed of real fragments that archaeologists have been studying for many years in Fig. \ref{fig:repair_puzzle}.
On this small puzzle, archaeologists annotated the patterns which we used as features for our players in a game without rotation. It is seen that the law of good continuation seems to be the wisest choice, as shown by the solution that our approach obtains, in Fig. \ref{fig:our_repair}.
This result confirms that our algorithm is a first step towards the solution of complex real-world puzzles, whereas the method by Derech et al. \cite{DERECH2021108065} for archaeological puzzles struggles due to the low discrimination power of color and shape in this case study.

\subsubsection{Limitations}
The challenging datasets highlight some limitations of our algorithm. 
The discretization of the space used to create the partial payoff relative transformation is a strong bottleneck when scaling to larger puzzles, making it a computationally expensive process.
Moreover, the variable density of the lines and the presence of empty regions mislead our algorithm in some of the map puzzles, alternating success and failure cases. These limitations will be tackled in future works.

\vspace{-0.2cm}
\section{Conclusions}
\vspace{-0.2cm}

Motivated by the observation that in real-world applications color and shape features used by existing jigsaw puzzle-solving algorithms are often unusable, in this paper we proposed a new challenging version of the problem which relies solely on the presence of linear geometrical patterns. To address this problem, we developed a principled game-theoretic framework which allows one to solve puzzles with pieces of arbitrary shape, size and orientation. Experimental results show the inadequacy of standard techniques as well as the relative 
advantages of our game-theoretic formulation. 

Future work will focus on extending the proposed ideas to more complex curve clues and on integrating good continuation features with standard shape and color information.



\subsubsection*{Acknowledgements}
This project has received funding from the European Union’s Horizon 2020 research and innovation programme under grant agreement No 964854. 

\bibliographystyle{splncs04}
\bibliography{main}
\end{document}